\documentstyle[11pt,newpasp,twoside,epsf]{article}
\def\edcomment#1{\iffalse\marginpar{\raggedright\sl#1\/}\else\relax\fi}
\reversemarginpar

\begin{document}
\title{{\it Chandra} Spectroscopy and Mass Estimation of 
the Lensing Cluster of Galaxies CL0024+17}
 \author{Naomi Ota}
\affil{Tokyo Metropolitan University, 1-1 Minami-osawa, 
Hachiouji, Tokyo 192-0397, Japan}
\author{Makoto Hattori, Etienne Pointecouteau}
\affil{Tohoku University, Aoba Aramaki, Sendai 980-8578, Japan}
\author{Kazuhisa Mitsuda}
\affil{ISAS, 3-1-1 Yoshinodai, Sagamihara, Kanagawa 229-8510, Japan}

\begin{abstract}
We present the X-ray analysis and the mass estimation of the lensing
cluster of galaxies CL0024+17 with {\it Chandra}. We found that the
temperature profile is consistent with being isothermal and the
average X-ray temperature is $4.47^{+0.83}_{-0.54}$ keV.  The X-ray
surface brightness profile is represented by the sum of emissions
associated with the central three bright elliptical galaxies and the
emission from intracluster medium (ICM) which can be well described by
a spherical $\beta$-model. Assuming the ICM to be in the
hydrostatic equilibrium, we estimated the X-ray mass and found it is
significantly smaller than the strong lensing mass by a factor of 3.
\end{abstract}

\section{Introduction}

CL0024+17 is one of the most extensively studied lensing clusters of
galaxies, located at $z=0.395$. Since the discovery of the multiply
lensed arc system, several authors modeled the matter distributions in
the cluster. Tyson, Kochanski, \& Dell'Antonio (1998) constructed a
very detailed mass map and suggested that the dark matter profile has
a soft core. Broadhurst et al. (2000) measured the arc redshift to
be 1.675 and also built a lens model in a simplified manner.

On the other hand, the X-ray emitting gas is an excellent tracer of
the dark matter potential. Soucail et al. (2000) performed a combined
analysis of the {\it ROSAT} and {\it ASCA} data and estimated the
cluster mass within the arc radius (the X-ray mass, hereafter). They
found that there is about a factor of $\sim3$ discrepancy between the
X-ray mass and the strong lensing mass (Tyson et al. 1998; Broadhurst
et al. 2000). Because the {\it ROSAT} HRI image suggested the
elongated gas distribution, they considered that the discrepancy may
be caused by the irregular mass distribution.

However, there were still large measurement uncertainties in both the
X-ray temperature and the image morphology. It was mainly because of
the heavy contamination from the bright seyfert galaxy. Thus for the
cluster mass estimation the temperature determination is crucial. In
this paper, we report on the accurate measurements of the temperature
and the morphology with {\it Chandra}, from which we discuss whether
there is an inevitable mass discrepancy between the X-ray and the
strong lensing. We use $H_0=50$ km/s/Mpc and $\Omega_0=1$. $1\arcmin =
383\,h_{50}^{-1}$ kpc at $z=0.395$.

\section{Observation}

We observed CL0024+17 with the {\it Chandra} ACIS-S detector on
September 20, 2000. The net exposure time is 37121 sec. In Figure 1a,
we show the ACIS-S3 image. The strongest X-ray peak is at (00:26:36.0,
+17:09:45.9)$_{\rm J2000}$ and the extended emission is detected out
to $\sim2\arcmin$ in radius. The point sources detected in the field
were removed in the following analysis.

\section{Spectral analysis}

We extracted the cluster spectrum from a circular region of $r
=1\arcmin.5$, centered at the X-ray peak (the dashed circle in Figure
1a). The background was estimated from the $2\arcmin.5 < r <
3\arcmin.2$ ring region. We fitted the spectrum to the MEKAL
thin-thermal plasma model with the Galactic absorption (Figure 2a) and
determined the temperature to be $kT = 4.47^{+0.83}_{-0.54}$ keV (90\%
error).  This is consistent with our previous result with {\it ASCA}
(Soucail et al. 2000; Ota \& Mitsuda 2002). We detected the strong
redshifted Fe-K line from the cluster for the first time. The iron
abundance is $0.76^{+0.37}_{-0.31}$ solar.

In order to investigate the radial temperature profile, we accumulated
spectra from four ring regions with various radii and fitted them with
the MEKAL model. The radius ranges were chosen so that the each
spectrum contains more than 400 photons. We found that there is not
any meaningful temperature variation against radius (Figure 2b); the
gas is consistent with being isothermal.

\section{Image analysis}
\subsection{X-ray surface brightness and galaxy distribution}

Though the original ACIS CCD has a pixel size of $0\arcsec.5$, we
rebinned the image by a factor of 4. We restrict the energy range to
0.5 -- 5 keV in the image analysis. We find that there is the second
X-ray peak at (00:26:35.1,+17:09:38.0)$_{\rm J2000}$. From comparison
with the galaxy catalog by Czoske et al. (2001), we recognized that
the three central bright elliptical galaxies are located at the
positions consistent with the fist and second X-ray peaks (G1 and G2,
hereafter). Note that G1 contains two of the three elliptical
galaxies (Figure 1a).

\subsection{2-D surface brightness distribution}

In order to determine the X-ray emission profile of the ICM, we fitted
the 2-dimensional surface brightness distribution with a model
consisting of three $\beta$ profiles which we consider to represent
emissions from two elliptical-galaxy components and ICM component;
$S(r)=\Sigma_{i=1}^{3} S_i ( 1+(r/r_{c,i}))^{-3\beta_i+1/2}$. We
fitted the image of $3\arcmin.3\times3\arcmin.3$ region with the
maximum-likelihood method. The center positions of the two
elliptical-galaxy components were fixed at the G1 and G2 peaks,
respectively, while for the third component, which we consider
describes the ICM emission, the position was allowed to vary. The
results of the fits are shown in Table 1 and Figure 1b.  In order to
check the goodness of the fit, we rebinned the image into two single
dimensional profiles of two perpendicular directions and calculated
the $\chi^2$ values between the model and data profiles to find they
are enough small ($\chi^2< 112 $ for 99 degree of freedom). The
best-fit cluster center position is $80\,h_{50}^{-1}$kpc away from the
G1 peak. We consider that the emission of G1 and G2 can be attributed
to the elliptical galaxies because of the small luminosities.
Furthermore, we tested the significance of the ellipticity of the
cluster image and found it is not significant ($\epsilon < 0.2$).

\vspace{-0.5cm}

{\small 
\begin{table}[thb]
\begin{center}
\caption{Results of the 2-D image fitting with the three $\beta$-models}
\begin{tabular}{lllll}\\\hline\hline
Model & Center position & $\beta$ & $r_c$ & $L_{\rm X,bol}$ \\
component& RA,Dec in J2000 &  & $h_{50}^{-1}$ kpc & erg/s \\\hline
G1 	& 00:26:36.0,+17:09:45.9 (F)& 1 (F) & $52^{+11}_{-9}$ & $5.5\times 10^{43}$\\
G2 	& 00:26:35.1,+17:09:38.0 (F)& 1 (F) & 10 (F) & $3\times 10^{42}$\\
Cluster & 00:26:35.6,+17:09:35.2$^{\dagger}$ & $0.71^{+0.07}_{-0.06}$& $210^{+33}_{-30}$& $4.5\times 10^{44}$\\\hline
\end{tabular}
\end{center}
(F) Fixed parameters. $^{\dagger}$ The 90\% errors are $\pm1\arcsec.3$ 
for RA and $\pm1\arcsec.5$ for Dec.
\end{table}
}

\section{Mass estimation and comparison}

From the spectral and spatial analysis mentioned above, we found that
the gas is isothermal and can be described with the spherical
$\beta$-model. Assuming the gas is hydrostatic, we obtained the
projected X-ray mass within the arc radius to be $M_{X,\beta}(<r_{\rm
arc}=220\,{\rm kpc}) = 0.85^{+0.12}_{-0.09}\times
10^{14}h_{50}^{-1}\,{\rm M_{\sun}}$. On the other hand, the strong
lensing mass was estimated to be $M_{\rm lens} (<r_{\rm arc})=(3.117
\pm0.004)\times 10^{14} h_{50}^{-1}\,{\rm M_{\sun}}$ by Tyson et
al. (1998). Therefore the discrepancy of a factor of 3 is
evident. This is consistent with our previous result (Soucail et
al. 2000). On the other hand, the X-ray surface brightness profile of
the NFW potential is similar to that of the $\beta$-model and can be
converted from the $\beta$-model parameters through the relations of
$r_s=r_c/0.22$ and $B=15\beta$ (Makino, Sasaki, \& Suto 1998). We thus
derived the gas profile for the NFW case based on the results of
$\beta$-model fitting and estimated the X-ray mass to be $M_{\rm X,
NFW}(<r_{\rm arc})=0.75^{+0.11}_{-0.05}\times10^{14}h_{50}^{-1}\,{\rm
M_{\sun}}$. Thus the discrepancy still remains even in this case.

Since the gas is isothermal and spherical, we consider that the gas is
relaxed in the cluster potential and the hydrostatic equilibrium is a
good approximation in the X-ray mass estimation. This suggests that
the lens mass is significantly overestimated. Czoske et al. (2001,
2002) measured the redshift distribution of galaxies in the direction
of CL0024+17 and revealed the presence of foreground and background
groups of galaxies. This may possibly enhance the strong lensing
mass. Thus updated lens modeling of the cluster is urged.

\begin{figure}[thb]
\plotone{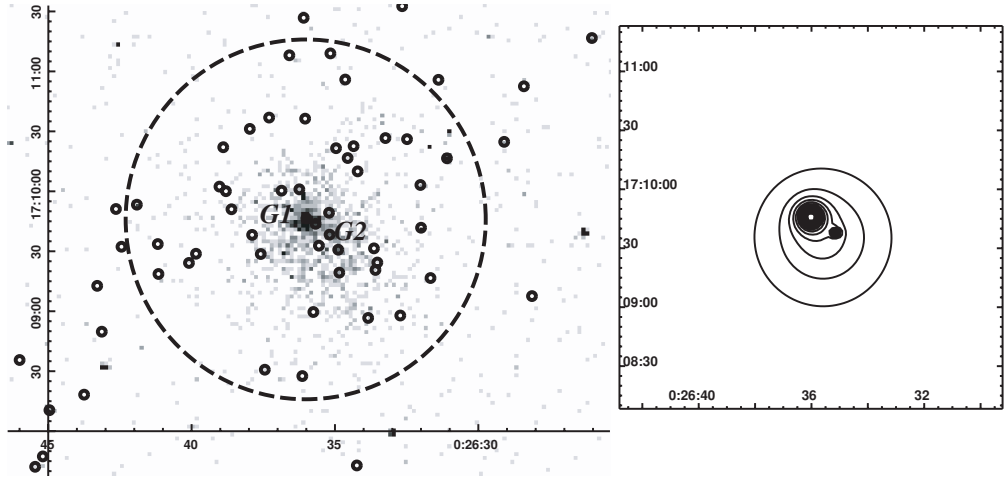}
\caption{\small (a) {\it Chandra} ACIS-S3 image of CL0024+17 
in the $0.5-5$ keV. 
The first and second X-ray peaks are labeled as G1 and G2. 
The small circles are the positions of the galaxies with 
$0.38 <z< 0.41$(Czoske et al. 2001). 
(b) Contours of the best-fit 2D image 
of the three $\beta$-models.}
\end{figure}

\begin{figure}[thb]
\plottwo{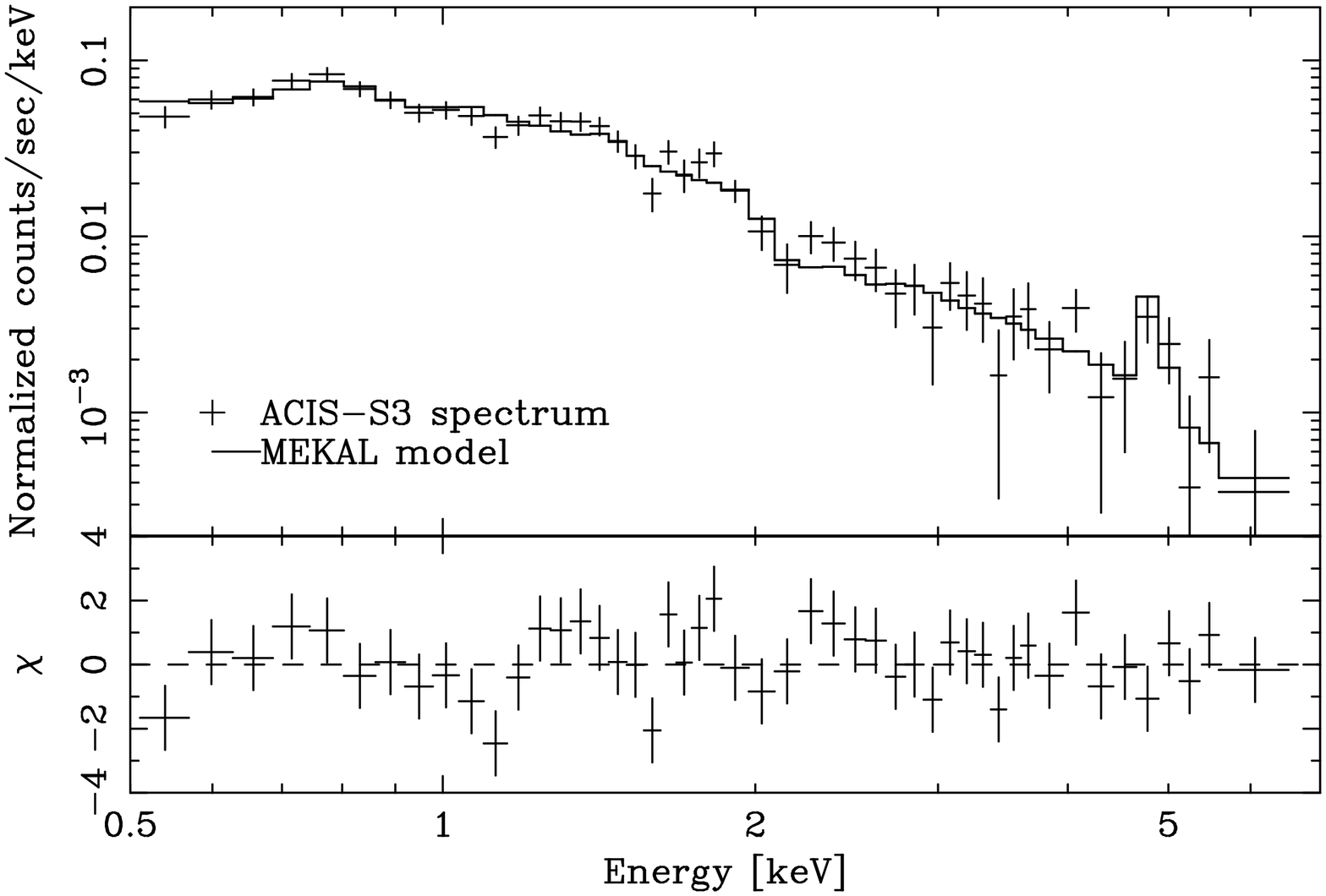}{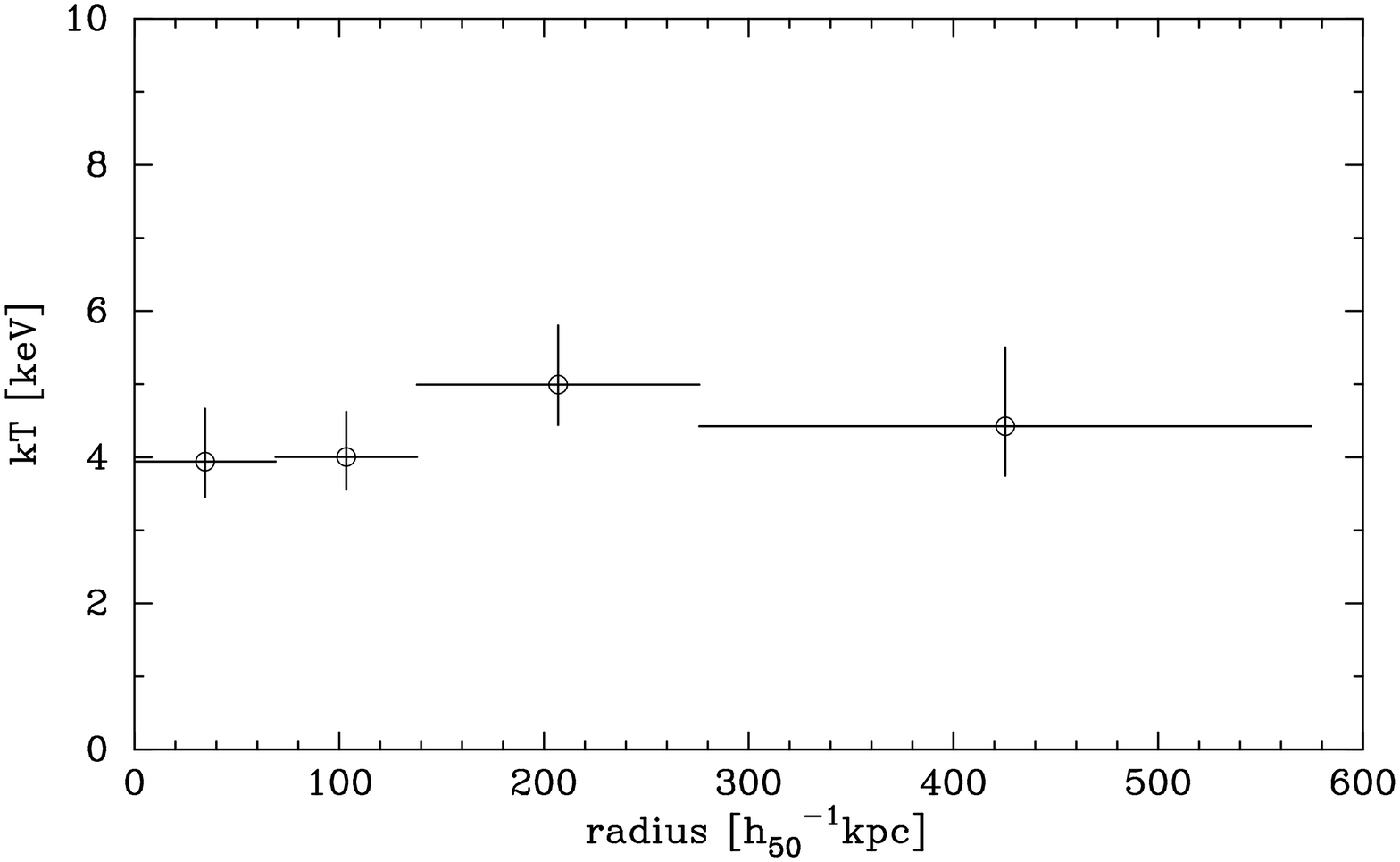}
\caption{\small (a) {\it Chandra} average spectrum of CL0024+17 fitted with the MEKAL model and (b) radial temperature profile.}
\end{figure}

\end{document}